\documentclass{article}
\usepackage{arxiv}
 \usepackage[american]{babel}
 \usepackage{graphicx}
\usepackage{commath}
\usepackage{booktabs} 
\usepackage[table]{xcolor}
\usepackage{siunitx} 
\usepackage{amssymb}

\usepackage{csquotes}
\usepackage{textcomp} 
\usepackage{float} 
\usepackage[style=ieee]{biblatex}
\usepackage{xcolor}
\usepackage[utf8]{inputenc}
\usepackage{hyperref}
\colorlet{LinkColor}{red}
\hypersetup{ 
	colorlinks=true,
	linkbordercolor =white, 
	urlcolor=gray, 
	filecolor=LinkColor,
	citebordercolor=LinkColor,
	citecolor=LinkColor,	linkcolor=LinkColor,  
	pdftitle=\@title
}
\usepackage[activate={true,nocompatibility},final,tracking=true,kerning=true,spacing=true,factor=1100,stretch=10,shrink=10]{microtype}
\let\fullcite\textcite
\begin{document}

\title{Heart Rate Variability as a Predictive Biomarker of Thermal Comfort}
\author{ \href{https://orcid.org/0000-0002-9128-7080}{\includegraphics[scale=0.06]{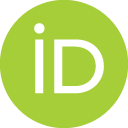}\hspace{1mm}Kizito N.~Nkurikiyeyezu, Yuta Suzuki, Guillaume F.~Lopez} \\
	Wearable Information Lab\\
	Aoyama Gakuin University\\
	4 Chome-4-25 Shibuya, Shibuya City, Tokyo 150-8366 \\
	\texttt{	\url{http://www.wil.it.aoyama.ac.jp/}} \\
}
\date{February 21, 2017}
\maketitle

\begin{abstract}
Thermal comfort is an assessment of one's satisfaction with the surroundings; yet, most mechanisms that are used to provide  thermal comfort are based on approaches that preclude physiological, psychological  and personal psychophysics that are precursors to thermal comfort. This leads to many people feeling either cold or hot in an environment that was supposed to be thermally comfortable to most users. To address this problem, this paper proposes to use heart rate variability (HRV) as an alternative indicator of thermal comfort status. Since HRV is linked to homeostasis, we conjectured that people's thermal comfort could be more accurately estimated based on their heart rate variability (HRV). To test our hypothesis, we analyzed statistical, spectral, and nonlinear HRV indice of 17 human subjects doing light office work in a cold, a neutral, and a hot environment. The resulting HRV indice were used as inputs to machine learning classification algorithms. We observed that HRV is distinctively different depending on the thermal environment and that it is possible to reliably predict each subject's thermal state (cold, neutral, and hot) with up to a 93.7\% accuracy. The result of this study suggests that it could be possible to design automatic real-time thermal comfort controllers based on people's HRV.
\end{abstract}
\section{Introduction}
\label{intro}
\par
Human thermal comfort is largely regulated using models based on static heat transfer equations. P.O. Fanger's Predicted Mean Vote/Predicted Percentage of Dissatisfied (PMV/PPD) model \cite{Fanger1970} is a noteworthy example of such models. These static thermal models, however, have well-documented limitations \cite{VanHoof2008}. For example, the PMV/PPD model is based on laboratory experiments on adults in highly controlled thermal chambers for a relatively long period. Thus, the model  intrinsically assumes fit adults who have achieved a steady state with the surrounding environment; thus, it should not be applied to any other group of people without amendments. Also, it assumes prior knowledge of people's clothing and their activity level. However, people's clothing and activities vary from one person to another and depending on the season of the year. Furthermore, it requires rigid and narrow comfort zones, which are never met in practice \cite{DeDearR;Brager1998a,VanHoof2008}. Finally, static models prescribe a fixed year-round set temperature point for all occupants and ignore any seasonal psychological adaptation on thermal comfort \cite{Lin2011}. Adaptive models have been introduced for an efficient thermal comfort provision. Contrary to the former, they recognize the influence of outdoor settings on how humans adapt to the environment \cite{DeDearR;Brager1998a} and are based on field studies \cite{DeDear2013b}. Moreover, because people prefer a broad range of thermal comfort zones \cite{Humphreys2007, Nastase2016, DeDearR;Brager1998a}, adaptive models' thermal ranges are not as limited as that of the static models. Recently, partly due to an increased awareness of global warming and the importance of environmental conservation, new innovative thermal comfort delivery mechanisms have been proposed. This includes personally controllable air conditionings, chilled beams, mixed-mode ventilation systems, and a plethora of other exotic methods as recapitulated in \fullcite{DeDear2013b}'s systematic review article. These contemporary systems provide an uneven thermal exposure resulting in a better energy efficiency.
\par
Despite the significant improvement in thermal comfort provision, experts in the fields acknowledge that current thermal comfort regulation mechanisms need comprehensive improvement\cite{OleFanger2001, Croitoru2015, DeDear2011a,Mahdavi1996a, Fountain1996, Brager2015, Nicol2017}. For example, despite their high energy consumption, the state of the art thermal regulation mechanism are expected to satisfy a mere 80\% of the building occupants \cite{ASHRAE2013}. Unfortunately, even this scant performance is not achieved in practice. In a survey of 215 buildings in North America and Finland, \fullcite{Huizenga2006} found that only 11\% of the buildings achieved the 80\% satisfaction rate. On average, $ 59 \pm 16\% $ satisfaction rate was achieved. This is obviously unacceptable as de Dear gloomily decried:
\begin{quote}
	\blockquote[{\cite[p.~115]{DeDear2011a}}]
	{If the very best that can be achieved in an isothermal, cool, dry and still indoor climate is  \enquote*{neutral} or \enquote*{acceptable} for little more than 80\% of a building's occupants at any one time, then the standards that have been set to date leave much to be desired}
\end{quote}
This dearth of thermal comfort delivery is further exposed by an \fullcite{InternationalFacilityManagementAssociationIFMA2009}'s survey on complaints in buildings. The survey concluded that many building occupants complain about too hot (91\% of the case) and too cold (94\%) temperatures. Surprisingly, complaints about other nuisances such as indoor air quality, noise level, limited office space are not as common. \fullcite{Fountain1996} make a convincing case for why this lack of thermal comfort is expected. They give an example of two people wearing a typical business attire and show that there is no PMV-based set-point temperature that would provide a satisfying thermal comfort for both of them. This is not too surprising since current thermal comfort provision mechanisms aim to provide a satisfying thermal comfort for an \enquote*{average} building occupant. However, since the standard deviation of thermal comfort expectations of a large group of people is high \cite{Ye2006}, in practice, an \enquote*{\textit{average thermal comfort}} is thermally dissatisfying to many people \cite{Fountain1996}. Moreover, most of these thermal comfort provision systems are based on heat transfer and energy conservation principles; thus, innately reflect only the influence of the environment on the person's thermal comfort. However, they neither relate  to the complexity of human thermo-regulation nor to the adequacy of the provided thermal comfort. For practical reasons, most thermal comfort provision mechanisms assume an all-embracing thermal comfort, i.e. that all people of all gender and age, with different psychological and physiological conditions, have the same thermal comfort expectations. In doing so, they ignore important personal psychophysics that are known to affect the perception of thermal comfort \cite{DeDearR;Brager1998a,Karjalainen2007, Karjalainen2012, Natsume1992a, Brager1998}. Another concern is that these systems make  a tacit assumption that people crave for thermal neutrality. A thermal neutral environment is described as an environment in which people do not wish to be warmer or cooler \cite{Mahdavi1996a}. In reality, however, people tend to prefer non-neutral conditions \cite{Humphreys2007,Nastase2016, Brager1998, Jones2002}. Ergo, achieving an outright thermal neutrality seem to be meretricious as it does not necessarily reflect neither thermal satisfaction nor thermal pleasure \cite{DeDear2011a}. In their articles, de Dear and his coauthors \cite{DeDear2009,DeDear2011a, Parkinson2016e,Parkinson2016d,Parkinson2016b,Parkinson2015a} highlighted numerous shortcomings of the one-size-fits-all static thermal settings advocated by most thermal regulation mechanisms. Instead, they proposed \textit{alliesthesia} which is a non-steady state thermal pleasure achieved by varying thermal sensation. It is based on the observation that someone in warm conditions is thermally pleased by cold temperature while the opposite is true for a cold person.
\par
\begin{figure}[!htb]
	\centering
	\includegraphics[width=1\linewidth]{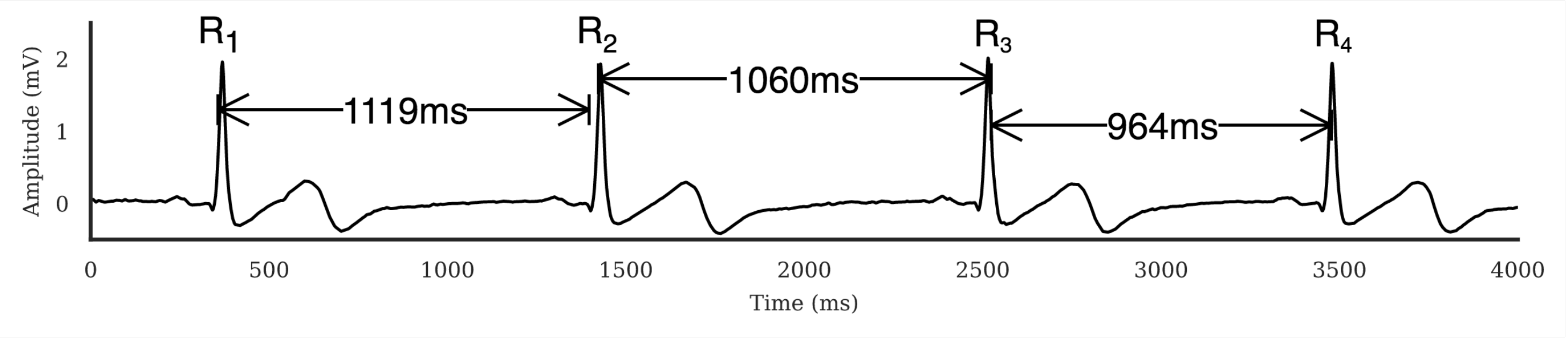}
	\caption{Heart Rate Variability: A time variation of consecutive heart beats (RR intervals)}
	\label{fig:fig1}
\end{figure}

 In this paper, we propose to use heart rate variability (HRV) to discretely classify people's thermal comfort state. HRV reflects the time variation between two successive heartbeats. Heartbeats intervals (also referred to as the R-R intervals) are not periodic. Rather, the duration between two consecutive R-R intervals varies from one heart beat to another (Figure \ref{fig:fig1}). However, this variation is not random; instead, it changes depending on complex extrinsic protocols imposed on the heart \cite{Acharya2006a}. Equally important, HRV is linked to homeostasis \cite{Riganello2012} i.e. the human body's ability to uphold optimum conditions despite external stimuli. For humans, the core temperature is regulated by the brain's hypothalamus that control various mechanism to increase or reduce energy production in order to restore the core temperature \cite{Hammel1968}. Homeostasis is a result of complementary actions of the parasympathetic nervous system (PNS) and the sympathetic nervous system (SNS). The two systems have an antagonistic effect on the heart rate: the SNS increases the heart rate while the PNS has an adverse effect \cite{Robinson1966, TaskForceoftheEuropeanSocietyofCardiologyandtheNorthAmericanSocietyofPacingandElectrophysiology1996}.
Since thermal comfort is, by definition, a personal subjective assessment of the satisfaction of the mind with the thermal environment \cite{ASHRAE2013} and given that thermal changes in environment affect homeostasis \cite{Riganello2012}, which in turn affect HRV \cite{Berntson1997}, our hypothesis  is that thermal comfort state could be more accurately predicted based on the persons HRV. The main long-term objective our research is to assess the possibility to fully automate indoor air conditioning based on people's HRV. Unlike existing thermal comfort estimation models, using HRV would allow a personalized real-time thermal comfort regulation. 
\par A thorough review of the relevant literature revealed that there is a need for such a study as existing research have limited scopes. For instance, in his various studies, Liu and his colleagues \cite{Liu2008a, Yao2009b} suggested that the sympathetic activity changes noticeably at different temperatures and concluded that heart rate variability may be used as an indicator for thermal comfort. These studies were however only limited to LF/HF, an HRV index whose origin and interpretation is ill-understood and a moot point \cite{Billman2013, Milicevic2005}.Another significant research was conducted at Carnegie Mellon University. This research, however, only focused on the heart rate and skin temperature \cite{Choi2010b, Choi2012b}. HRV studies have also been used in sport and exercise medicine. For example, \cite{Almeida2016} demonstrated that HRV indice return to baseline after 60 minutes of post-exercise recovery. This duration is reduced to just 10 to 20 minutes when Cold Water Immersion (CWI) is used. 
\section{Material and Method}
\label{metho}
\subsection{Experiment}
\par Seventeen male university students (age:$ 22.35 \pm 1.08 $ years, weight:$59.75 \pm 7.78$ kg, BMI:$ 20.07 \pm2.26 $) voluntarily participated in the study. All subjects were healthy and did not suffer from any cardiovascular or respiratory illness. Prior to the experiment, the subjects were requested to abstain from eating, drinking coffee, smoking or doing extensive physical exercise at least two hours before the start of experiment. The experimental protocols were approved by the ethics committee at our university. Subjects wore light clothes (long-sleeved T-shirt and long pants) and were placed inside a climate chamber. During the experiment, the subjects were seated and asked to type randomly selected news articles to simulate office work. The experiments were conducted in three climate chambers whose thermal settings (Table \ref{table:chambersettings}) are similar to a cold, a neutral and a hot thermal sensation on the  PMV index scale adjusted for cooling effect of elevated air speed \cite{ASHRAE2013, Berkeley2009}. The subjects' electrocardiograms (ECG) were recorded at a 1000Hz sampling rate and saved for later analysis. The experiment lasted for about 30 minutes. Right after the experiment, the subjects were requested to rate their thermal comfort sensation on Visual Analog Scale (VAS) consisting of a 10 equal intervals line with numerical ranges from 0 (representing the lowest sensation) on the left side to 10 (the highest sensation) on the right side of the line. The subjects were asked to mark a position on the line that best reflect their thermal comfort level, a lower position number indicating a thermal dissatisfaction with the environment.  
\begin{table}[!htb]
  \centering
  \caption{Climate chamber thermal settings}
  \begin{tabular}{rccc}
    & \multicolumn{3}{c}{} \\
    \midrule
    & Cool  & Neutral & Hot \\
    \midrule
    Activity level & 1     & 1     & 1 \\
    Clothing level & 1     & 1     & 1 \\
    Air temperature(\textdegree{}C) & 18.0  & 24.0  & 30.0 \\
    Radiant temperature(\textdegree{}C) & 18.0  & 24.0  & 30.0 \\
    Air speed (m/s) & 0.3   & 0.3   & 0.3 \\
    Humidity (\%RH) & 50.0  & 65.0  & 80.0 \\
    PMV index$^{\dagger }$ & -1.79 & -0.03  & +1.87 \\
    \bottomrule
    \multicolumn{4}{c}{ $^{\dagger }$PMV adjusted for the cooling effect of an elevated air speed} \\
  \end{tabular}%
  \label{table:chambersettings}%
\end{table}%

\subsection{HRV Calculation}
We developed an \emph{ad hot} HRV analysis software. The software enables, \emph{inter alia}, the extraction of temporal, spectral and nonlinear HRV indice either from a raw ECG signal or from an R-R interval signal. The extraction of R-R interval signal from a raw ECG is based on the \fullcite{Pan1985} algorithm and the HRV indice are calculated according to standards and algorithms proposed by the 
\textit{\fullcite{TaskForceoftheEuropeanSocietyofCardiologyandtheNorthAmericanSocietyofPacingandElectrophysiology1996}} and its subsequent revision by the \textit{e-Cardiology ESC Working Group and the European Heart Rhythm Association} (2015). In this study, heuristically-selected statistical, spectral, and nonlinear HRV analysis methods were used to calculate various HRV indice summarized in Table \ref{table:listofhrvindices}. 
\begin{table*}[!htb]
  \centering
  \caption{Short description of the selected HRV indice}
    \begin{tabular}{rl}
    \toprule
          & \multicolumn{1}{c}{Time domain HRV Indice} \\
\cmidrule{2-2}   $  Mean RR  $& Average of all RR intervals \\
    $ RMSSD $ & Square root of the mean of the sum of difference of successive RR intervals \\
    $ SDSD $  & Standard deviation of difference between adjacent RR intervals \\
    $ pNNx $  & Percentage of RR pairs that differ by x milliseconds in the entire recording \\
          &  \\
          & \multicolumn{1}{c}{Spectral HRV Indice} \\
\cmidrule{2-2}    TP    & Total spectral power (0-0.4Hz) \\
    $ VLF $   & Spectral power in very low range frequencies (0.003-0.04Hz) \\
    $ LF  $   & Spectral power in low range frequencies (0.04-0.15 Hz) \\
   $  HF $    & Spectral power in high range frequencies (0.15 -  Hz) \\
    $ LF/HF $ & Ration between LF and HF power \\
          &  \\
          & \multicolumn{1}{c}{Non Linear HRV Indice} \\
\cmidrule{2-2}          &  \\
    $ DFA(\alpha1) $ & Short-term fluctuations of detrended fluctuation analysis  \\
    $ DFA(\alpha2) $ & Long-term fluctuations of detrended fluctuation analysis \\
    $ SD1, SD2 $ & Short and long-term variability of the Poincar\'e plot \\
    $ SampEn $ & Sample Entropy -- A measure of complexity \\
    \bottomrule
    \end{tabular}%
  \label{table:listofhrvindices}%
\end{table*}%
 
Statistical time domain HRV indice describe the beat-to-beat variability using statistical techniques. They are easy to compute and self-explanatory. In this group, the RMSSD and pNNx indice are of a great interest to the HRV research community. The RMSSD represents the square root of the mean of the sum of difference of successive R-R intervals. It is calculated (Equation \ref{equ:rmssd}) as
\begin{equation}\label{equ:rmssd}
RMMSD =\sqrt{\frac{1}{N-1}\sum_{i=1}^{N-1} (RR_{i+1}-RR_{i})^{2}}
\end{equation}
The pNNx, with $ x>0 $, is a family of statistical HRV measures of cardiac vagal tone modulation. It indicates the percentage of R-R consecutive pairs that differ by x milliseconds (Equation \ref{equ:pnnx} )
\begin{equation}\label{equ:pnnx}
pNNx =\frac{1}{N-1}\sum_{i=1}^{N}\big ( \left|R_{i}-R_{i+1}\right|>x \big )
\end{equation}
Unlike descriptive statistical methods, spectral HRV analysis methods decompose the heart beat variability into its fundamental frequency components; thus, provide a greater understanding of the fluctuation of the heartbeats. Spectra methods provide an assessment of the interplay between the autonomic nervous system and the heart \cite{Berntson1997}. There exist many approaches to estimating HRV spectra components. Notably, the Fast Fourier Transform (FFT) and the autoregressive (AR) modeling-based techniques are commonly used. However, the two methods assume stationary input signals (in other words, signals whose mean and standard deviation are time invariant). Therefore, for HRV studies, they should be used with caution as the R-R wave signal is not stationary and is naturally irregularly sampled. In this study, the Lomb-Scargle Periodogram(LSP) \cite{Lomb1976} was instead preferred. Contrary to most other spectral estimation methods, LSP is suitable to non-stationary and irregularly sampled signals such the R-R wave and it is recommended as the most suitable method for HRV spectra analysis \cite{Laguna1998, Moody1993, Fonseca2013}. A normalized LSP power  of a signal $ X $ of $ N $ data points sampled at time $t_i$ and with $\overline{X}$as its mean is estimated as (Equation \ref{equ:lomb})
\begin{equation}\label{equ:lomb}
P(w) =\begin{Bmatrix}
\frac{\big[\sum_{i=1}^{N} (X_{i}-\overline{X}cos[w(t_i -\tau)])\big]^{2}}{2\sigma ^{2} \sum_{i=1}^{N} cos^{2}[w(t_i -\tau)]}\enspace+ \cdots\\
\enspace+ \cdots \frac{\big[\sum_{i=1}^{N} (X_{i}-\overline{X}sin[w(t_i -\tau)])\big]^{2}}{2\sigma ^{2} \sum_{i=1}^{N} sin^{2}[w(t_i -\tau)]}\\
\end{Bmatrix}
\end{equation}
with $w$ and $ \sigma^{2}$ the angular frequency and variance respectively, and the time constant $\tau$ defined as in equation \ref{equ:tau}.
\begin{equation}\label{equ:tau}
\tau = \frac{1}{2w}tan^{-1} \Bigg(\frac{\sum_{i=1}^{N} sin(2wt_i)}{\sum_{i=1}^{N} cos(2wt_i)}\Bigg)
\end{equation}
In HRV studies, four spectra bands are normally distinguished: the ultra-low frequency $(ULF[0 - 0.0033 Hz])$, the very low frequency $(VLF[0.0033Hz-0.04Hz])$, the low frequency $(LF[0.04Hz-0.15Hz])$, and the high frequencies $(HF[0.15-0.4Hz])$. The spectral power of each band is estimated by calculating the area under the spectral density function over its frequency band. Lastly, physiological processes are characterized by phenomena that cannot be measured neither by statistical nor spectral methods \cite{West2010, Sassi2015, Hardstone2012}. Therefore, non-linear HRV analysis methods have been used to account for any non-linear and multi-fractal dynamics in heart beats variation. A systematic review of such methods can be found in the recent literature \cite[e.g.][]{Sunkaria2011, Sassi2015}. In this study, three non-linear HRV indice were used: the Poincar\'e plot\cite{Piskorski2007, Brennan2001}, the sample entropy\cite{Lake2011, Richman2000} and the Detrended Fluctuation Analysis (DFA)
\cite{Peng1995b, Willson2002, Echeverria2003}. \fullcite{Maestri2007} have shown that these three non-linear HRV indice are the most reliable for HRV analysis. A Poincar\'e plot is a diagram in which an $ RR_i $ point is plotted against its adjacent $ RR_{i+1} $ point. Thus, for an R-R signal $ RR_1, RR_2, RR_3, …RR_N $, a Poincar\'e plot consists of points $ (X, Y) $  described (Figure \ref{equ:poincarepoints}) as,
\begin{align}
\label{equ:poincarepoints}
\begin{split}
X =RR_1, RR_2, RR_3, \cdots RR_{N-1},
\\
Y =RR_2, RR_3, RR_4, \cdots RR_{N}
\end{split}
\end{align}
A Poincar\'e plot has two main descriptors: A short $ (SD1) $ and long $ (SD2) $  term descriptor defined in equations \ref{equ:sd1} and \ref{equ:sd2}.
\begin{equation}\label{equ:sd1}
SD1=\sqrt{var(x_1)}
\end{equation}
\begin{equation}\label{equ:sd2}
SD2=\sqrt{var(x_2)}
\end{equation}
with $ var(x) $ the variance of x and $ x_1 and \ x_2$ defined as in equation \ref{equ:x1x2}

\begin{equation}\label{equ:x1x2}
\left(\begin{array}{c}x_1\\ x_2\end{array}\right) =
\begin{pmatrix}cos(\pi/4) & -sin(\pi/4) \\ sin(\pi/4) & cos(\pi/4)
\end{pmatrix}\left(\begin{array}{c}X\\ Y\end{array}\right)
\end{equation}
The $ SD1 $ and $ SD2 $ are respectively reported to represent the short term and long term variability. The $ SD1 $ is believed to indicate the parasympathetic activity while the $ SD2 $ index reflects the overall HRV variability\cite{Acharya2006a}. The sample entropy is another important method used to quantify the predictability of a signal. The entropy itself is a measure of an average information content in a signal\cite{Shannon1948}. The sample entropy uses conditional probability to measure the likelihood that two sequences of length $ m $ (that match each other with a tolerance $ r $) will still match if a new sample is added \cite{Humeau-Heurtier2015}. The sample entropy of a signal $ X $ is defined as
\begin{equation}\label{equ:sampleEn}
SampEn(m,r) =-\ln\bigg[\frac{A^{m}(r)}{B^{m}(r)}\bigg]
\end{equation}
where$  A^{m}(r) $ is the probability that the two sequences will match if a new sample is added to the sequence (thus, the signal is $ m+1 $ points) and $  B^{m}(r) $, the probability that the two sequences (of length $ m $) will match. Thus, the sample entropy measures the variability of signal when an extra sample is added. A low entropy signal is more predictable than a higher entropy signal. The sample entropy, unlike some other entropy measures, is much robust against noisy biomedical data such the R-R signal \cite{Richman2000}. It is important to note however that its algorithm is influenced by its parameters $ m $ and $ r $. Accordingly, the two parameters should be carefully selected\cite{Yentes2013}. In this study, our algorithm uses $ m=2 $ and $ r=0.2SDRR $. The Detrended Fluctuation Analysis (DFA) is yet another nonlinear HRV analysis method. It repeatedly assesses the signal at different times. Thus, in the case of physiology signal, DFA allows distinguishing short-term perturbations $ DFA(\alpha1) $ from perturbation dues to long lasting perturbations $ DFA(\alpha1) $ which are due to e.g. the body's internal functionality. In our study, the algorithm is implemented according to an algorithm described by \fullcite{Voss2012}. The algorithm consists of the following four major steps. First, the $ R-R $ signal is integrated (Equations \ref{equ:dfa1}).
\begin{equation}\label{equ:dfa1}
y_{int}(k) =\sum_{i=1}^{k} (RR_{i}-\overline{RR}), k =1,2, \cdots N
\end{equation}
In the above equation,$ \overline{RR} $ is the average of the signal. Second, the integrated signal is split into $ N $ non-overlapping equal length segments of length $ n $. Third, each segment is \enquote*{detrended} by removing a polynomial fit calculated using a least-square regression. Lastly, the Root Mean Square (RMS) fluctuation value $ F(n) $ of the signal is computed (Equation \ref{equ:dfa2})
\begin{equation}\label{equ:dfa2}
F(n) =\frac{1}{N}\sum_{k=1}^{N} \bigg(y_{int}(k)-y_{fit}(k)\bigg)^{2}
\end{equation}
In this study, the DFA's short term exponent $ DFA(\alpha1) $  was computed for a range of n=4 to 16, while the long-term exponent $ DFA(\alpha2) $ was computed for $ n =16 $ to 64 as recommended by \fullcite{Peng1995b}.

\subsection{Statistical Analysis}
The normal distribution of each HRV indice was tested using a Levene's test. The test revealed that most HRV indice were not normally distributed; thus, a Mann-Whitney-Wilcoxon test for two independent samples was used to test the statistical significance of HRV parameters in a different environment. This test is a non-parametric test that estimates the probability that a random sample from one group is less than or greater than another randomly selected sample from another group; thus, this test is used to compare two independent groups of data.
\subsection{Machine Learning Classification}
Supervised Machine Learning (ML) classification algorithms were used to build models that could be used to predict the comfort status of a person. In this study, Logistic Regression (LR), Linear Discriminant Analysis (LDA), K-Nearest Neighbors (KNN), Decision Tress (DT), Na\"ive Bayes (NB), Support Vector Machine (SVM), Random Forest (RF), Multi-Layer Perceptron(MLP),  Adaptive Boosting (ADABOOST), and Quadratic Discriminant Analysis (QDA) classifiers were trained to predict if a subject is in a cold, neutral, or hot environment based on the subject's HRV indice. \fullcite{Kotsiantis2006} provide a concise review and algorithmic implementations of the aforementioned ML classifiers. Because there exist many HRV indice, some of which may contain redundant information, we used a Randomized Logistic Regression(RLR) \cite{Meinshausen2010} to select the most important HRV indice for classifiers. In a nutshell, the RLR algorithm fits a randomly selected subset of the data; then, it combines and analyzes the results from different HRV indice. The HRV indice deemed less significant in classifying the data are rejected while the important one are retained. The RLR method has a twofold advantage: It increases the accuracy of the model and reduces the complexity and computation requirements of the model. The classification accuracy of each ML classifier was estimated using a 10-fold stratified cross-validation.
\subsection{HRV Indices Computation}
In this study, we were interested in understanding how thermal environment affect the evolution heartbeat. Thus, to compute the HRV indice listed in Table \ref{table:listofhrvindices}, for each subject, the HRV indice were first extracted from a five-minutes-long ECG window segment. Then, the window segment is shifted by approximately 15 seconds and new HRV indice are calculated. This process is repeated until the end of the entire ECG recording. In this paper, since different HRV indice have different numerical ranges, the result is reported on a normalized scale between 0 and 1.
\section{Results}
\begin{figure}[!htb]
	\centering
	\includegraphics[width=1\linewidth]{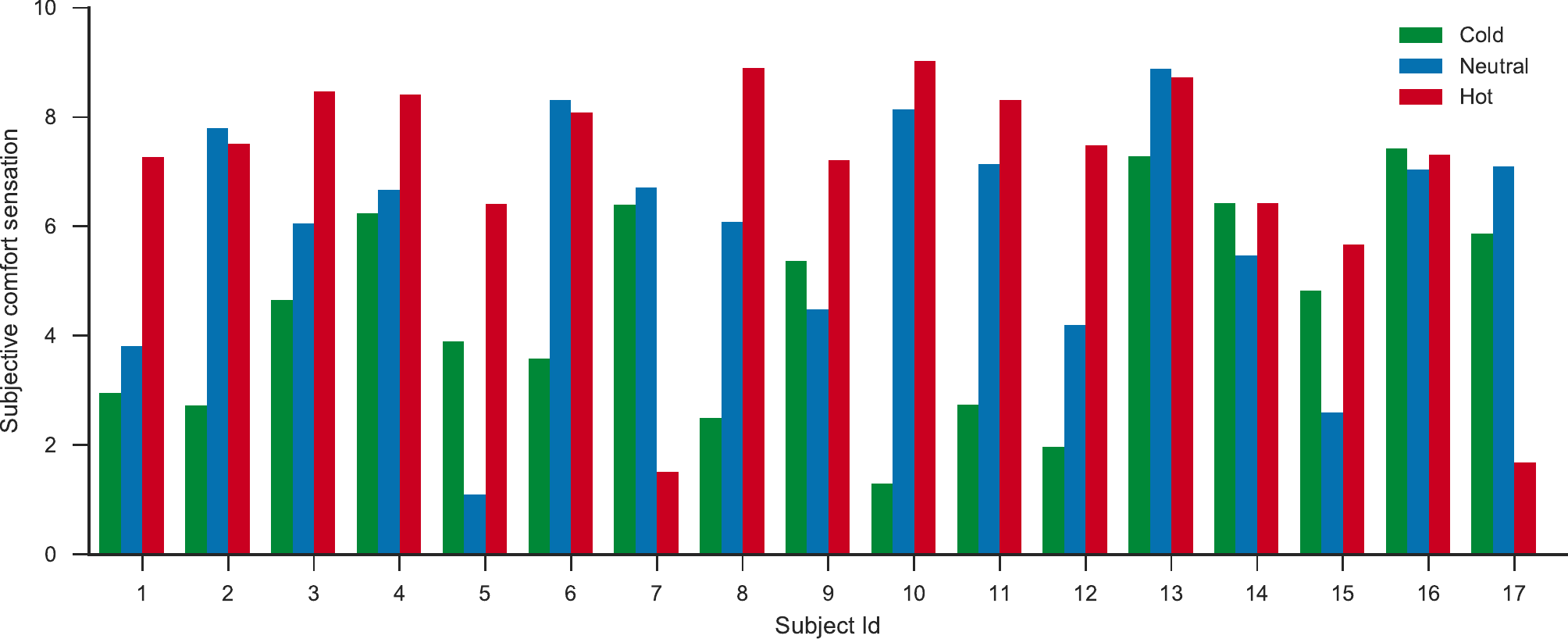}
	\caption{\textbf{Self-reported thermal sensation}\newline
	The subjects rated their thermal sensation on a visual analogue scale (VAS).Although the thermal environments were the same, the subjects had a different thermal sensation for the same environment}
	\label{fig:vasresults}
\end{figure}

\begin{figure}[!htb]
	\centering
	\includegraphics[width=1\linewidth]{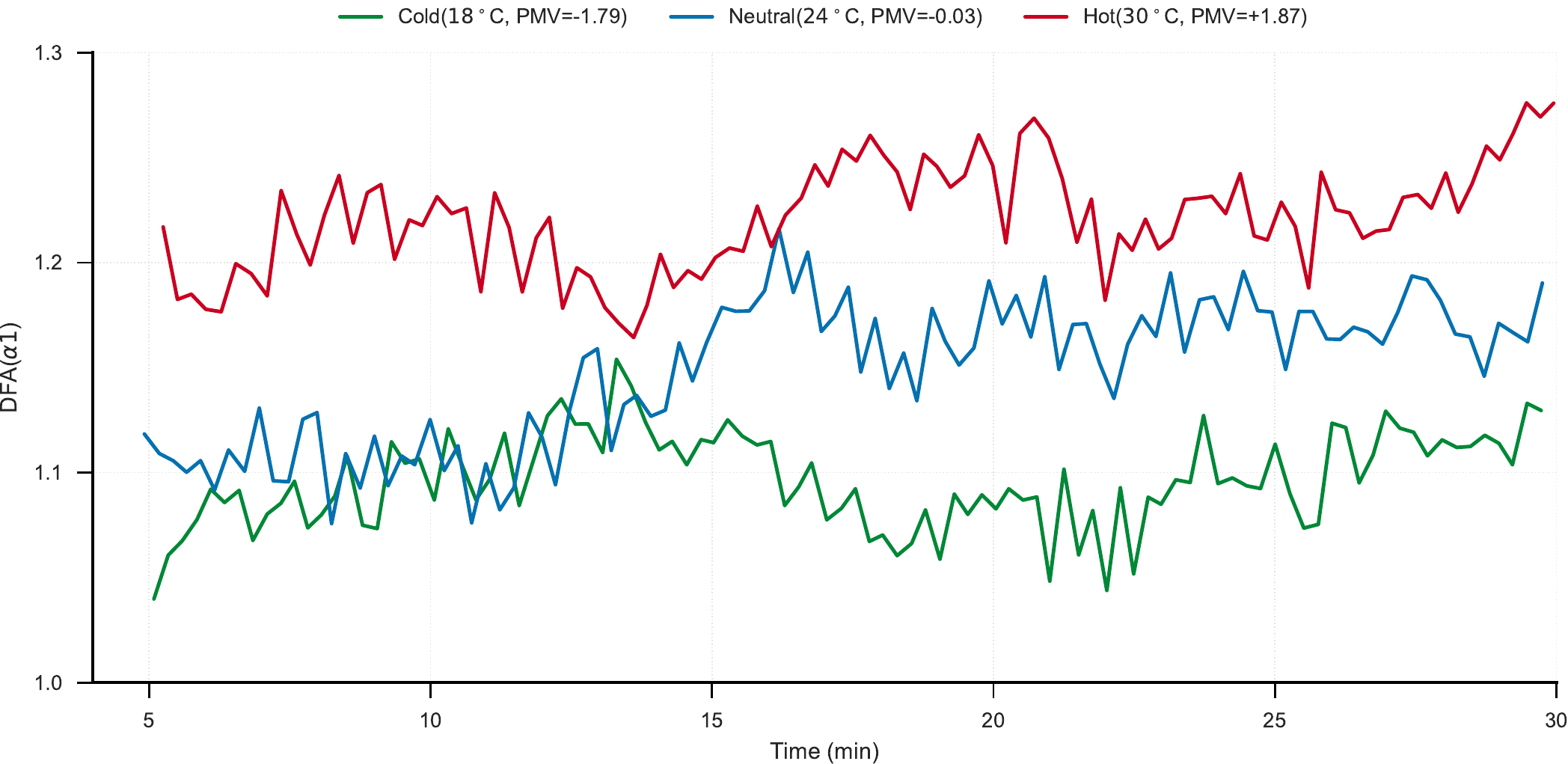}
	\caption{Mean $  DFA (\alpha1) $ indicating that the $  DFA (\alpha1) $ is consistently highest in the hot environment and mostly lowest in the cold environment}
	\label{fig:highestinhot}
\end{figure}

\begin{figure}[!htb]
	\centering
	\includegraphics[width=1\linewidth]{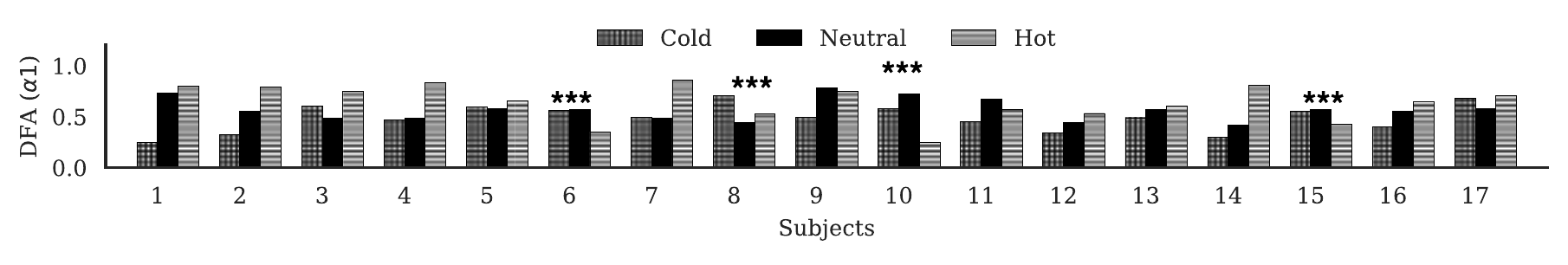}
	\caption{Inter-subject comparison of a normalized  indicating that, for most subjects, the $ DFA(\alpha1) $ is highest in the hot environment and lowest in the cold environment. The tree star sign (***) indicates irregularity in this trend, i.e. where the $ DFA(\alpha1) $ was higher in the cold than in the hot environment.}
	\label{fig:highestinhothistogram}
\end{figure}

\begin{figure}[!htb]
	\centering
	\includegraphics[width=1\linewidth]{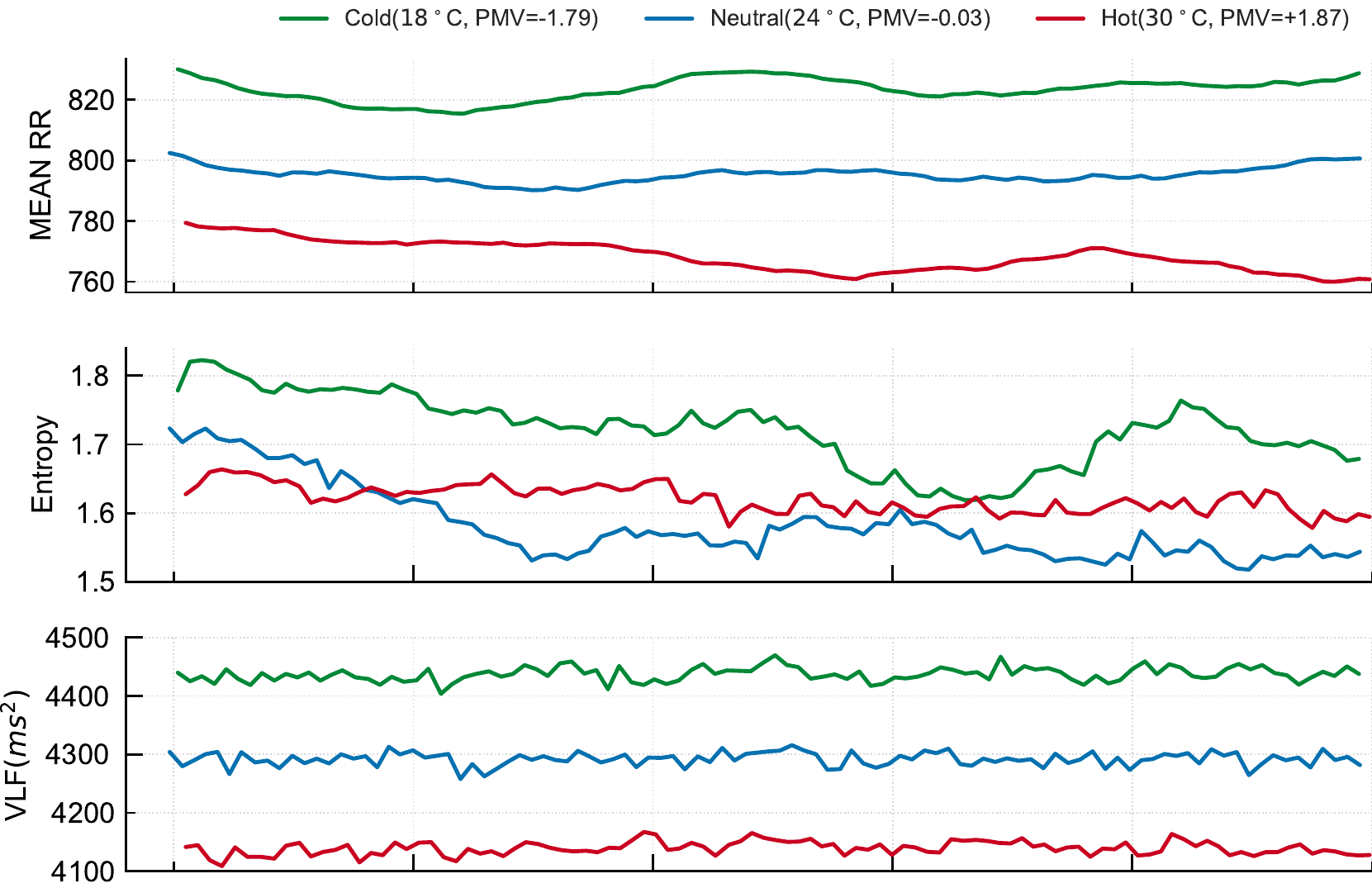}
	\caption{The mean of HRV indice that are highest in the cold environment.}
	\label{fig:highestincold}
\end{figure}

\begin{figure}[!htb]
	\centering
	\includegraphics[width=1\linewidth]{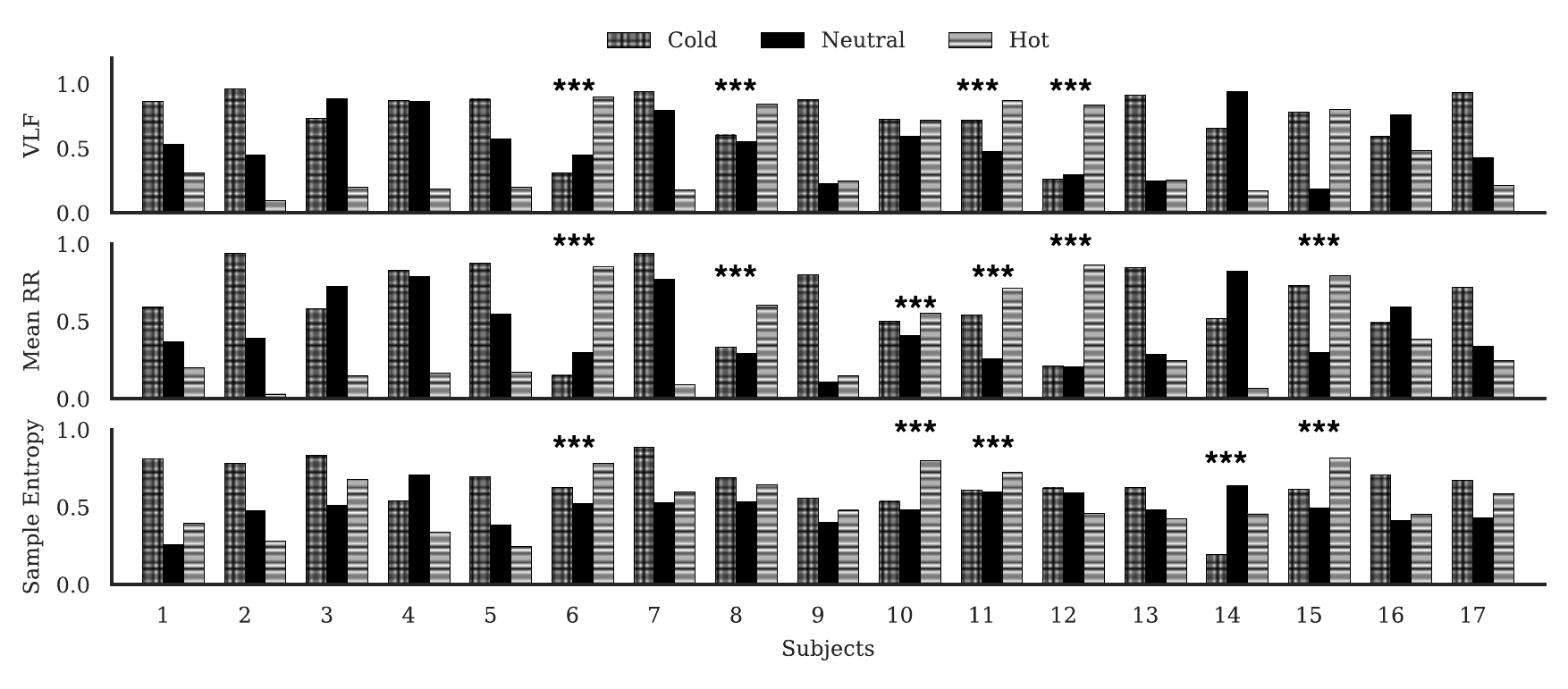}
	\caption{Inter-subject comparison of normalized HRV indice highest in the cold environment. The three stars (***) indicates irregularity in this trend, i.e. where the HRV indice were higher in the hot than in the cold environment.}
	\label{fig:highestincoldhistogram}
\end{figure}

\begin{figure*} [!htb]
	\centering
	\includegraphics[width=1\linewidth]{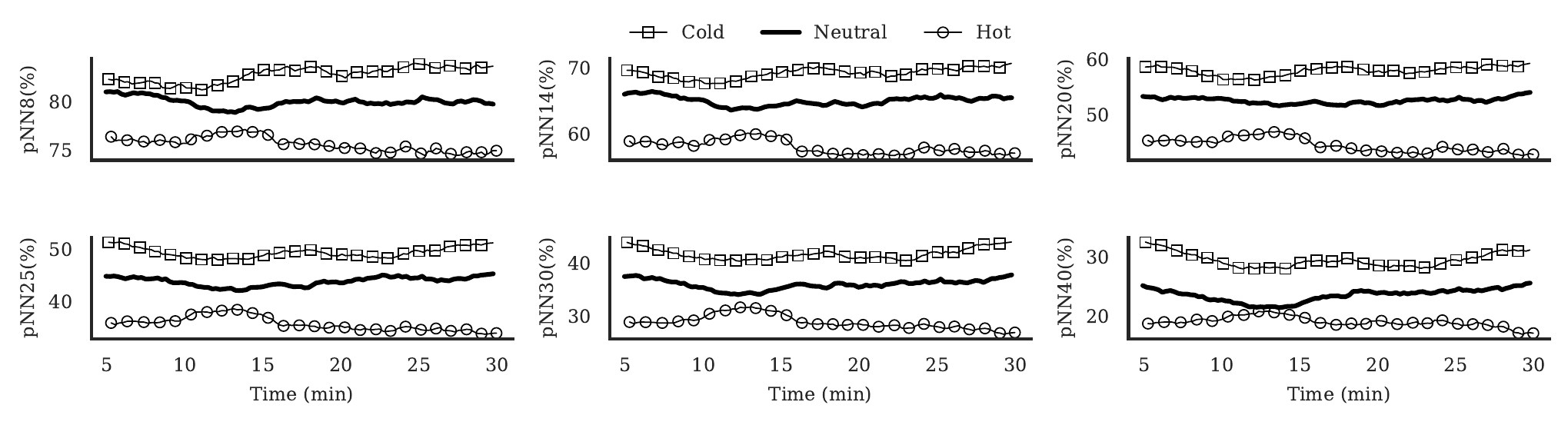}
	\caption{Time variation of some pNNx HRV indice indicating that pNNx is highest in the cold environment and lowest in the hot environment}
	\label{fig:pnnvstime}
\end{figure*}

\begin{figure}[!htb]
	\centering
	\includegraphics[width=1\linewidth]{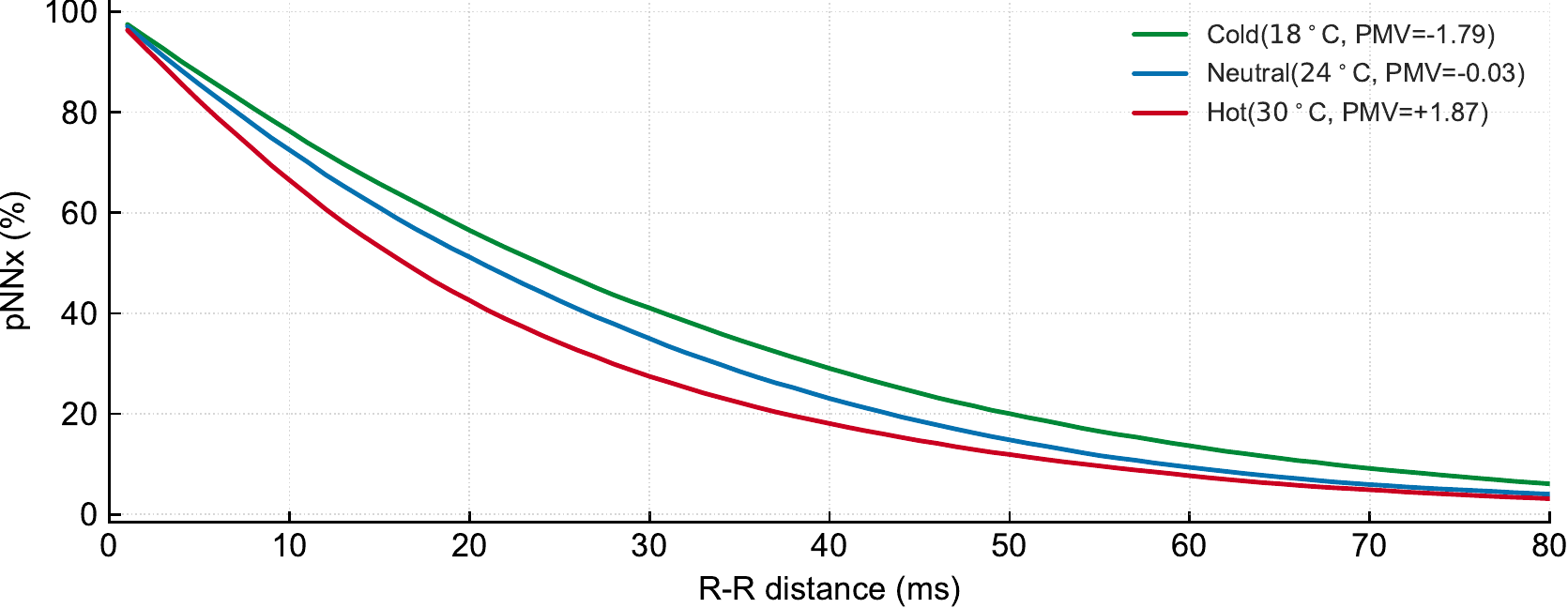}
	\caption{pNNx variation with an increase in an R-R increment x. Note that the separation between the hot and cold environment is highest between 20 and 40 milliseconds and gradually decreases outside this boundary.}
	\label{fig:pnnvsrrincrement}
\end{figure}

\begin{figure*}[!htb]
	\centering
	\includegraphics[width=1\linewidth]{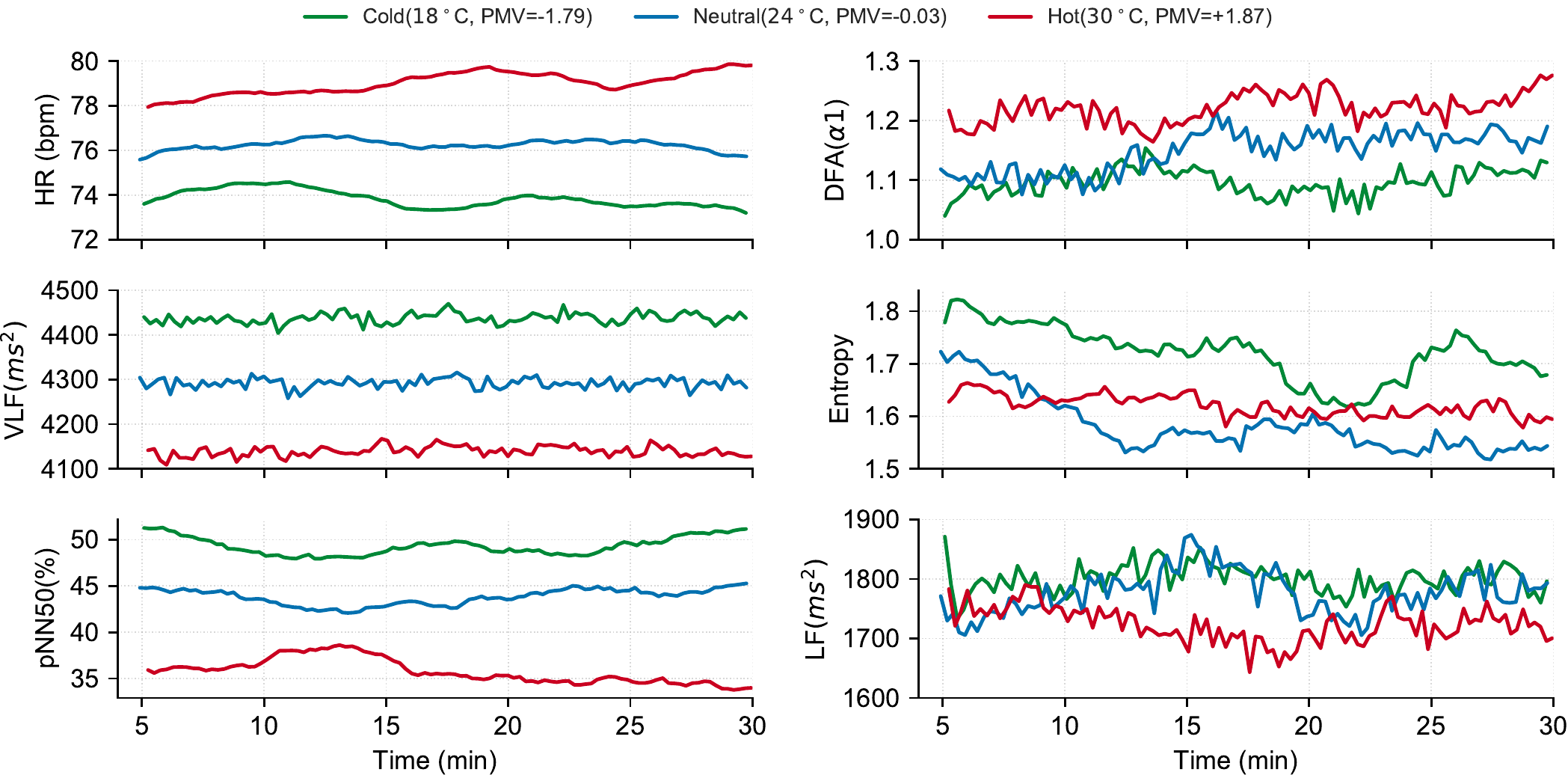}
	\caption{\textbf{Influence of thermal comfort on heart rate variability}\newline Each line of the plots represent the mean (17 subjects, 38200 samples) HRV fluctuation and show that HRV is distinctively influenced by the thermal environment ($P < 0.01$)}
	\label{fig:irregularhrvindices}
\end{figure*}

In general, subjects reported feeling cool or cold in the cold chamber, comfortable in the neutral thermal chamber and warm or hot in a hot thermal chamber. However, as shown in figure \ref{fig:vasresults}, different subjects reported a different level of thermal sensations for the same thermal environment.
A comparison of HRV in the three different environments reveals that HRV is distinctly different from one environment to another. A thorough analysis of the different HRV indices uncovers four important observations. First, in general, the $ DFA (\alpha1) $ is disproportionately highest in the hot environment and mostly lowest in the cold environment as illustrated in Figure \ref{fig:highestinhot} and summarized in table \ref{tab:hrvindicesvalues}. 
\begin{table*}[!htb]
	\centering
	\caption{Comparison of HRV indice values in different thermal settings}
	\begin{tabular}{rccc}
		\toprule
		& $ Cold (mean \pm std) $ & $ Neutral (mean \pm std) $ & $ Hot (mean \pm std) $ \\
		\cmidrule{2-4}    
		 MEAN RR(ms) & 824.47 $ \pm $ 91.42 & 795.13$ \pm $ 79.55 & 768.69$ \pm $ 83.52 \\
		SDRR(ms) & 54.25 $ \pm $ 19.33 & 55.56$ \pm $ 23.51 & 44.34$ \pm $ 13.32 \\
		
		RMSSD(ms) & 43.37 $ \pm $ 20.57 & 42.73$ \pm $ 30.74 & 31.00$ \pm $ 14.15 \\
		SDSD(ms) & 43.36 $ \pm $ 20.57 & 42.73 $ \pm $ 30.74 & 31.00$ \pm $ 14.15 \\
		
		pNN25($ \% $) & 49.49 $ \pm $ 18.11 & 43.77$ \pm $ 16.95 & 35.84$ \pm $ 20.79 \\
		VLF($ ms^{2} $) & 4442.82 $ \pm $ 483.55 & 4291.14$ \pm $ 418.26 & 4141.95$ \pm $ 442.64 \\
		
		LF($ ms^{2} $) & 1800.04 $ \pm $ 206.71 & 1777.18$ \pm $ 203.12 & 1723.08$ \pm $ 213.18 \\
		HF($ ms^{2} $) & 2554.58 $ \pm $ 491.47 & 2607.48$ \pm $ 545.67 & 2482.16$ \pm $ 528.45 \\
		
		TP($ ms^{2} $) & 8834.73 $ \pm $ 987.00 & 8713.29$ \pm $ 925.75 & 8383.47$ \pm $ 928.97 \\
		LF/HF & 0.72$ \pm $ 0.09 & 0.70$ \pm $ 0.11 & 0.71$ \pm $ 0.11 \\
		
		Sample Entropy & 1.72$ \pm $ 0.22 & 1.58$ \pm $ 0.30 & 1.62$ \pm $ 0.24 \\
		DFA ($ \alpha1 $) & 1.10$ \pm $ 0.21 & 1.15$ \pm $ 0.24 & 1.22$ \pm $ 0.23 \\
		
		SD1(ms) & 76.47$ \pm $ 27.28 & 78.33$ \pm $ 33.19 & 62.50$ \pm $ 18.79 \\
		SD2(ms) & 69.71$ \pm $ 24.53 & 71.53$ \pm $ 27.50 & 58.30$ \pm $ 17.18 \\
		SD1/SD2 & 1.10$ \pm $ 0.06 & 1.09$ \pm $ 0.07 & 1.07$ \pm $ 0.05 \\
		\bottomrule
	\end{tabular}%
	\label{tab:hrvindicesvalues}%
\end{table*}%
However, this trend is not unanimous to all subjects. A few subjects violate this general trend as indicated in figure \ref{fig:highestinhothistogram}. 
Second, the Mean RR, the VLF and the sample entropy HRV indice are noticeably highest in the cold environment and lowest in the hot environment  as illustrated in
figure \ref{fig:highestincold} and recapitulated in Table \ref{tab:hrvindicesvalues}.
However, unlike the $ DFA(\alpha1) $, the separation between the three environments is clearer and persisted during the whole duration of the experiment. As with the $ DFA(\alpha1) $, some subjects were exceptional to this trend (Figure \ref{fig:highestincoldhistogram}).
Third, the pNNx family of HRV indice are also highest in the cold environment and lowest in the hot environment (Figure \ref{fig:pnnvstime}).
Moreover, as shown in Figure \ref{fig:pnnvsrrincrement}, this difference is much more pronounced for an R-R increment roughly between 20 and 40 milliseconds and gradually decreases outside this boundary.
Fourth, for other HRV indice, the separation between the hot and cold environment is not as clear over the whole duration of the experiment. Instead, as shown in Figure \ref{fig:irregularhrvindices}, there is no clear distinction between the three thermal environments as the plots get mixed up with time. This is the case for spectral indice (TP, LF, LF/HF and HF), the Poincare plots based indice, and some statistical HRV indice such as the RMSSD, the SDRR, and the SDSD.
A Randomized Logistic Regression (RLR) feature selection algorithm was applied to obtain the most relevant features for thermal comfort prediction. The algorithm selected the following six HRV indice: Mean RR, the RMSSD, the SDSD, the pNN25, the VLF and the sample entropy. These HRV indice were used to make ML models that were afterward used to predict the thermal environment of a subject given his HRV indice. The highest accuracy was achieved by a Support Vector Machine (SVM) classifier at 93.7\% (Table \ref{table:classifiersaccuracy}).
\begin{table*}[!htb]
  \centering
  \caption{Cross-validation classification accuracy for each subject}
    \begin{tabular}{ccccccccccc}
     \toprule
          & \multicolumn{10}{c}{Machine Learning Classifier Accuracy (\%)} \\
\cmidrule{2-11}    Subject & LR    & LDA   & KNN   & DT    & NB    & SVM   & RF    & MLP   & ADBOOST & QDA \\
\cmidrule{2-11}
    1     & 93.18 & 96.89 & 93.86 & 91.83 & 94.76 & 96.59 & 95.26 & 96.3  & 85.62 & 95.98 \\
    2     & 100   & 100   & 100   & 100   & 100   & 100   & 100   & 100   & 100   & 100 \\
    3     & 90.72 & 94.87 & 92.36 & 92.25 & 86.76 & 96.79 & 96.39 & 94.88 & 93.13 & 89.75 \\
    4     & 79.66 & 81.19 & 82.18 & 86    & 77.52 & 90.5  & 88.85 & 88.55 & 86.52 & 85.17 \\
    5     & 100   & 100   & 100   & 99.02 & 100   & 100   & 99.33 & 100   & 99.02 & 99.68 \\
    6     & 90.6  & 99.02 & 98.69 & 97.69 & 99.02 & 99.35 & 95.83 & 99.67 & 96.45 & 97.7 \\
    7     & 91.79 & 99.67 & 99.67 & 100   & 99.33 & 100   & 100   & 100   & 100   & 99.33 \\
    8     & 80.49 & 79.89 & 84.19 & 75.78 & 71.23 & 80.66 & 84.62 & 81.43 & 71.85 & 77.42 \\
    9     & 77.38 & 71.85 & 79.16 & 82.64 & 75.4  & 81.55 & 80.3  & 75.63 & 75.01 & 78.06 \\
    10    & 92.68 & 95.42 & 91.69 & 91.36 & 89    & 96.09 & 93.91 & 95.45 & 89.7  & 98.36 \\
    11    & 70.25 & 74.55 & 74.74 & 77.49 & 66.68 & 82.99 & 84.39 & 81.62 & 66.24 & 74.63 \\
    12    & 95.71 & 95.38 & 93.85 & 96.25 & 93.74 & 96.28 & 93.52 & 95.3  & 96.55 & 96.89 \\
    13    & 74.38 & 73.64 & 80.72 & 82.27 & 75.03 & 83.98 & 77.7  & 80.42 & 74.52 & 82.49 \\
    14    & 99.67 & 100   & 98.47 & 98    & 100   & 100   & 100   & 99.69 & 100   & 100 \\
    15    & 99.7  & 100   & 100   & 100   & 100   & 100   & 100   & 99.09 & 100   & 100 \\
    16    & 88.39 & 85.81 & 91.71 & 86.68 & 83.3  & 93.64 & 89.01 & 90.01 & 68.17 & 93.39 \\
    17    & 87.61 & 97.78 & 93.54 & 92.63 & 92.84 & 95.27 & 96.82 & 97.09 & 97.78 & 95.83 \\
    \midrule
    Average  & 88.95 & 90.94 & 91.46 & 91.17 & 88.51 & 93.75 & 92.70 & 92.65 & 88.27 & 92.04 \\
    \bottomrule
    \end{tabular}%
  \label{table:classifiersaccuracy}%
\end{table*}%

\section{Discussion}
\label{discussion}
\par Despite advancement in thermal comfort alleviation technologies and the amount of energy used to provide it, most building occupants are not satisfied with the thermal environment in the building. A prominent survey \cite{InternationalFacilityManagementAssociationIFMA2009} brought to light this lack of comfort satisfaction and blames too hot and too cold temperatures as the main catalysts for the discomfort.This may be due to current thermal regulation systems that mostly ignore personal physiological and psychological precursor to thermal comfort. In this paper, we investigated the possibility to use HRV as an alternative indicator of thermal comfort. The study was conducted in three artificial thermal environments: cold, neutral, and hot on a PMV scale. In general, most subjects reported feeling cool or cold in the cold chamber, comfortable in the neutral environment and warm or hot in a hot environment. However, the level of thermal sensation slightly fluctuated from one subject to another. While there is a possibility that this variation is due to a bias of self-report questionnaires, other researchers \cite{Humphreys2007, Nastase2016, Brager1998, Jones2002} believe that different people have different thermal preferences. This discrepancy is due in part to variations in physiological and physiological differences among people. It may also be a result of inter-subjects thermal perception variability. In all cases, however, no subject expressed being hot in a cold environment or being cold in a hot environment; thus, we concluded that the subjects felt cold in a cold environment and hot in a hot environment.
\par The study found that HRV evidently varies from one environment to another $(p < \num{1e-4})$. The short-term DFA coefficient is consistently highest in the hot environment and lowest in the cold environment. The DFA is a measure of self-affinity and long term fractal correlation in the signal\cite{Peng1995b}. In HRV analysis, DFA is known to provide unique information that cannot be obtained from classical HRV analysis methods\cite{Hardstone2012, Penzel2003} and is used to for example to distinguish healthy subjects from unhealthy ones\cite{Rodriguez2007}. It is worthy to note that, while there is a clear separation between the hot and cold environments, for the $ DFA(\alpha1) $, the split-up between the cold and neutral environment is not as strong (Figure \ref{fig:highestinhot}). 
\par Other indice such as Mean RR, the VLF, and the sample entropy are highest in the cold environment and lowest in the hot environment. The VLF and the sample entropy are particularly interesting for our study. Unlike other HRV indice thus far, VLF is a result of analyzing the spectral components of the variability of successive R-R intervals. Thus, it reflects best the action of the nervous system on the heart rate variability. Moreover, some studies have linked VLF to thermal regulation\cite{Fleisher1996, Thayer1997}; thus, it could be a good indicator of thermal sensation. As for the sample entropy, it is highest in the cold and hot environments and lowest in the neutral environment(Figure \ref{fig:highestincold}). The sample entropy is a measure of unpredictability. In our case, it seems that the sample entropy is highest in non-thermally comfortable environments (hot and cold) and lowest in a comfortable environment (neutral). This suggests, that in an absence of thermal comfort, the heart beats are more unpredictable (thus, the increase in entropy). This increase in sample entropy suggests that the heart's functionality is influenced by some new external influence, and in this study, that external influence can only be the change in thermal environment conditions. This seems to be further confirmed by analyzing the balance between the parasympathetic nervous system (PNS) and the sympathetic nervous system
(SNS) in the three thermal environments. The parasympathetic system is responsible for the  \enquote*{rest and digest} activities to restore the balance of the body systems. In case of any perceived stressful situation, these activities are momentary minimized yield to the \enquote*{fight or flight} sympathetic nervous system. As shown in Table \ref{tab:hrvindicesvalues}, HF is highest the neutral condition and is lowest in the two uncomfortable thermal conditions (cold and hot). Since HF is known as an indicator of the cardiac parasympathetic nerve activity, it is safe to conclude that, in non-thermal comfortable environment, the body increases its parasympathetic activity to the detriment of the parasympathetic activity. This results in the increase of the LF/HF ratio which is believed to correlated with the cardiac sympatho-vagal balance (Table \ref{tab:hrvindicesvalues}). The pNNx family of HRV measure is also steadily highest in the cold environment and lowest in the hot environment. As shown in figure \ref{fig:pnnvsrrincrement}, the different is largest for an R-R increment between 20 and 40 milliseconds. For our study, the pNNx is also of a special interest. Unlike other HRV indice, it is less computation intensive to calculate; thus, could be used to estimate thermal comfort using less powerful embedded micro-controllers. 
\par We then investigated the possibility to automatically predict if people feel cold or hot based on their HRV. Since thermal sensation is linked to homeostasis which in turn affect HRV, we posit that HRV could be used to predict when people are cold, comfort, or hot. We assumed that, since the subjects' self-report survey indicated that they were indeed cold in a cold environment and hot in a hot environment, our prediction based on their HRV would closely reflect their thermal sensation. Our machine learning-based classification model achieved an 93.7\% accuracy. Although the highest classification accuracies was achieved by a complex models, much simpler, and less computation intensive models such as the NB classifiers provide adequate accuracy(Table \ref{table:classifiersaccuracy}).The results of this study lead us to believe, as we initially surmised, that it is likely possible to design automated thermal controllers that predict people's comfort state based on their HRV.
\section{Conclusion}
\par In our quest to design a more efficient thermal comfort controller, we proposed HRV as a biomarker for thermal comfort. The following findings emerged from this study:
\begin{enumerate}
	\item HRV is distinctively altered depending on the thermal environment. Some HRV indice were highest in the hot environment (notably the $ DFA(\alpha1)) $ and others were highest in the cold environment (the VLF, mean RR, the sample entropy, and the pNNx). For these indice, the difference between the hot and the cold environment was steady throughout the duration of the experiment.
	\item The sample entropy was highest in the uncomfortable environments (the cold and the hot). Since a high entropy indicates an increase in unpredictability in a signal, this confirms our assumption that a change in thermal environment indirectly influences a detectable change in the heart's functionality as the person adapts to the thermal stress. This suggests that the heart beat is more regular and less complex in thermal comfortable environment and exhibits a more complex pattern in non-thermal comfortable environments.
	\item The cardiac sympatho-vagal balance increases in non-thermal comfortable environment (cold and hot) and decreases in comfortable (neutral) settings.
	\item The study pinpointed six HRV indice that are most significant in predicting the thermal state of the subjects. These are: The Mean RR, the RMSSD, the SDSD, the pNN25, the VLF and the sample entropy
	\item Using machine learning algorithms, it was possible to reliably predict each subject's thermal state (cold, neutral, and hot) with up to a 93.7\%accuracy.
\end{enumerate}
\section{Practical Implication}
\par The results of this study suggests that it could be possible to design automated real-time thermal comfort controlled based on people's HRV. While we recognized that this is not a panacea for all thermal comfort requirements, we believe that, since non-neutral thermal conditions do not result necessarily in thermal discomfort, it could be possible to design a computerized system that let the indoor temperature drift away from the thermal neutrality and adjust it only if people feel cold or hot. This approach would reduce the energy used for thermal comfort regulation while providing acce ptable thermal comfort. In their simulations, \cite{Hoyt2014} demonstrated that it could be possible to save 29\% of cooling energy and 27\% total HVAC energy by simply expending cooling setpoint by a mere \SI{2.8}{\celsius}. Likewise, lowering the heating setpoint by \SI{1.1}{\celsius} has a potential to save an average of 34\% of heating energy. Furthermore, due to the availability of non-contact ECG sensors and a proliferation of inexpensive wearable devices with built-in photoplethysmography (PPG) sensors (e.g. smart-watches and fitness trackers), this objective could be achieved in a non-invasive manner and at low cost. PPG is a reliable unobtrusive technique to monitor beat-to-beat heart rate variability via arterial oxygen saturation and is known as a good HRV surrogate  \cite{Selvaraj2008, Bolanos2006}.
\printbibliography
\end{document}